\documentclass[preprint2]{aastex}
\begin{document}
\title{{\itshape Chandra} Observations of High Mass Young Stellar Objects 
in the Monoceros~R2 molecular cloud}

\author{M. Kohno, K. Koyama}
\affil{Department of Physics, Graduate School of Science, Kyoto
University, Sakyo-Ku Kyoto, 606-8502}
\email{kohno@cr.scphys.kyoto-u.ac.jp, koyama@cr.scphys.kyoto-u.ac.jp}
\and
\author{K. Hamaguchi}
\affil{NASDA TKSC SURP, 2-1-1 Sengen, Tsukuba, Ibaraki 305-8505}
\email{kenji@oasis.tksc.nasda.go.jp}

\begin{abstract}

We observed the Monoceros~R2 molecular cloud with the ACIS-I array
onboard the {\itshape Chandra X-ray Observatory}. From the central
$3\farcm2 \times 3\farcm2$ region, we detect 154 sources above the
detection limit of $\sim 5\times 10^{-16}~{\rm ergs~s^{-1}~cm^{-2}}$
 with a 100~ks-exposure. 
About 85\% of the X-ray sources are identified with an infrared counterpart, 
including four high mass stars in zero age main sequence (ZAMS) and/or pre 
main sequence (PMS) phase. The X-ray spectra of the high mass ZAMS and PMS 
stars are represented with 
a thin thermal plasma model of a temperature above $\sim 2$~keV.
The X-rays are time-variable and exhibit  rapid flares. 
These high temperature plasma and  flaring activity 
are similar to those seen in low mass PMS stars and contrary to the behavior 
observed in high mass main sequence stars.
The X-ray luminosity increases as the intrinsic $K$-band flux increases. 
However, the X-ray luminosity 
saturates  at  a level of $\sim 10^{31}~{\rm ergs~s^{-1}}$.
We conclude that high mass  ZAMS and PMS emit X-rays, possibly due 
to the magnetic activity like those of low mass stars.
\end{abstract}


\section{INTRODUCTION} 

X-ray emission from normal  stars are generally attributed to  
either  magnetic activity, stellar winds or both.
Low mass young stellar objects (YSOs) emit X-rays with occasional flare 
activities.  The X-ray spectra are described with a thin thermal  plasma model of 
a temperature ranging from  1 to a few keV.  
These types of time variability and spectral shape are similar  to those of  the 
solar X-rays but with much larger luminosity, which lead to the general consensus that 
the X-ray origin is  due to the enhanced  solar-type magnetic activity; magnetic 
amplification and release of the field energy (Feigelson \& DeCampli 1981;
Montmerle et al.\ 1983).
The X-rays become less active as low mass stars evolve to the main
sequence (MS) stars. 

High mass MS stars lack the mechanism of magnetic field 
amplification, but they emit moderately variable X-rays.   
  The X-ray spectra are also due to a thin 
thermal plasma with the temperature (less than 1~keV) lower than those
of low mass stars.  
The  X-ray flux is approximately proportional to the bolometric luminosity,
or strength of the stellar wind  (Bergh\"ofer et al.\ 1997).
Thus the stellar wind may be involved in the  origin of X-rays from high mass stars; it may
be either the heated plasma
by the shock induced in the stellar wind (Lucy \& White 1980) or the
wind collision with nearby stars (Pittard \& Stevens 1997). 
It has also been suggested that X-rays can be produced by the interaction of a stellar
wind with a stable (i.e. dipolar like) magnetic field (e.g.\ Gagn\'e et al.\ 1997).

 X-ray observations of high mass YSOs are largely behind those of low mass stars and 
high mass MS stars, because of less samples of this class
due to their quick evolution and less population than  those of low mass stars.  
In fact, the 
distance to the nearest star forming region (SFR) of high mass stars, the Orion Nebula  is 
three times further than those of low mass SFRs.   
Since high mass YSOs  generally reside in the dense cores of giant molecular 
clouds (GMCs), soft X-rays should be strongly absorbed, hence hard X-ray imaging 
instruments are essentially required. 
The {\itshape Advanced Satellite of Cosmology and 
Astrophysics} ({\itshape ASCA}) has found  hard X-rays from the center of GMCs, the site of 
high mass star formation: the Orion region (Yamauchi et al.\ 1996),  W3
(Hofner \& Churchwell 1997), NGC6334 (Sekimoto et al.\ 2000), and Monoceros R2
(Hamaguchi, Tsuboi, \& Koyama 2000).  However, the limited spatial resolution of
$\sim 1'$, did not allow us to uniquely  resolve the X-rays from high mass YSOs.
Recently, {\itshape Chandra X-ray Observatory} observed the
Orion Nebula and confirmed hard X-rays from high mass 
stars, the trapezium stars and  ``Source~n'' in 
the early MS phase (Schulz et al.\ 2001; Garmire et al.\ 2000).  

The Monoceros R2 cloud (here, Mon~R2) is  a SFR at a distance of 
830~pc (Racine et al.\ 1968).  The central region has a shell-like complex (IR shell) 
of sub-mm and far infrared dust cores (IRS~1--3) with  $K$-band stars
(Aspin \& Walther 1990; Henning, Chini, \& Pfau 1992).  
The core masses are 30--150~${\rm M}_{\odot}$,  which are one order of magnitude 
larger than those of protostellar cores of low and medium mass star forming regions.
Thus the cloud cores (IRS~1--3) are good candidates for high mass star formation.  

IRS~1 is resolved into two IR stars, IRS~1SW and 1NE (Howard, Pipher,
\& Forrest 1994). Howard et al.\ (1994) identified IRS~1NE with optical
star ``B'' (Cohen \& Frogel 1977). 
Since there is an emission peak in the radio and IR band and the 
degree of polarization in the IR band is small
around IRS~1 (Yao et al.\ 1997),
the brighter  source IRS~1SW of bolometric luminosity $3.0 \times 10^3~L_\odot$
(Henning et al.\ 1992) is 
a B0 type star in zero age main sequence (ZAMS) and excites a compact 
\ion{H}{2} region inside the IR shell (Massi, Felli, \& Simon 1985). 

The IR source ${\rm a_S}$ is associated with a small IR nebulosity and has
a similar IR spectrum 
to that of IRS~1SW (Carpenter et al.\ 1997), hence would be the same
class, a high mass star of early B type. Since ${\rm a_S}$ has no
\ion{H}{2} region, it would be younger than the ZAMS star IRS~1SW. 

IRS~2 is an illuminating source of the IR shell. 
The infrared spectrum of IRS~2 (and also IRS~3, see the next paragraph)
shows deep absorption in the 
water-ice band (Smith, Sellgren, \& Tokunaga 1989), and the
spectropolarimetry of IRS~2 shows an excess polarization across the ice
band (Yao et al.\ 1997), which is the signature  of  a cluster
of several young embedded sources  
including  BN-like objects (Hough et al.\ 1996), indicating that it  is
younger than the ZAMS star IRS~1SW. However, no evidence for a molecular
outflow from IRS~2 have been reported. The bolometric luminosity of 
IRS~2 is $6.5 \times 10^3~L_\odot$ (Henning et al.\ 1992).
  
IRS~3, the brightest near- and mid-IR source in the Mon R2, has
a  bolometric luminosity of 
$1.3\times 10^4~L_\odot$, and is another active star forming 
site (Henning et al.\ 1992). The presence of ${\rm H}_{2}$O and OH
masers (Smits, Cohen, \& Hutawarakorn 1998) and
a compact molecular outflow indicate that IRS~3 is still in a phase
of dynamical mass accretion (Giannakopoulou et al.\ 1997).
IRS~3 has been resolved into two sources, IRS~3NE and
SW (Carpenter et al.\ 1997).

IRS~1, ${\rm a_S}$, IRS~2 and 3, thus comprise nice samples for the evolution of  
high mass stars from pre main sequence (PMS) to ZAMS. We therefore
performed  a deep {\itshape Chandra} observation  of  Mon~R2.  
This paper reports the results and discusses the X-ray evolution in the early 
phase of high mass stars.

\section{OBSERVATION AND DATA REDUCTION} 

We observed the Mon~R2 dark cloud on December 2--4, 2000 with the 
Advanced CCD Imaging Spectrometers (ACIS) 
onboard the {\itshape Chandra  X-ray Observatory} 
(Weisskopf et al.\ 2000).
We used the ACIS-I array consisting of four abutted X-ray CCDs, which covers
a full region of Mon~R2 and surroundings. 
Using the Level 2 processed events provided by the pipeline
processing at the Chandra X-ray Center, we selected 
the ASCA grades\footnote{see
http://asc.harvard.edu/udocs/docs/POG/MPOG/index.html} 0, 2, 3, 4 and
6, as X-ray events; the other events, which are due to charged
particles, hot and flickering pixels, are removed. 
The effective exposure is then about 100 ks.

\section{ANALYSIS AND RESULTS} 

\subsection{Source Detection and  Identification} 

The ACIS X-ray image of a  $3\farcm2 \times 3\farcm2$ region at Mon~R2
is given in Figure 1 with the blue 
and red colors representing  the hard (2.0--10.0~keV) and soft (0.5--2.0~keV) bands, 
respectively.  The CS $J = 5 \rightarrow 4$ line intensity map (Choi et al.\
2000) is overlaid on the X-ray image.
To search for X-ray sources in the region, 
we perform {\itshape wavdetect}\footnote{see
http://asc.harvard.edu/udocs/docs/swdocs/detect/html/}
in the total (0.5--10.0~keV), hard (2.0--10.0~keV) and soft
(0.5--2.0~keV) band images. We find 142 sources with a significance criterion at $10^{-6}$
and the wavelet scales ranging between 1 and 16 pixels. In addition, we manually inspect the
image and find 12 sources above 5 $\sigma$ confidence level.
The mean position error is $0\farcs15$.
X-ray events from each detected source are extracted within a radius of
$2\farcs5$, which is about 5 times the FWHM of 
the point spread function at the on-axis position. 
These circles  include about 90 \% of the total 
photons from the relevant sources, 
but include less than one  background event, hence we do not subtract the background.
In crowded regions, we extract events within a radius of $1\farcs5$ to
avoid contamination.

We search for a near-infrared counterpart from the deep imaging in the {\itshape J}, {\itshape H},
{\itshape K} and {\itshape nbL'} band by Carpenter et al.\ (1997), 
using IRCAM3 at UKIRT.
In their measurement, the typical uncertainty is about 0.05~${\rm mag}$ for all these
bands. 
Since the X-ray positions are systematically offset to the northeast, we
correct the X-ray positions by shifting $1\farcs50$ to the west and $0\farcs76$ 
to the south.
After adjusting the X-ray positions,
the offset between the X-ray source and its IR counterpart becomes 
$\sim 0\farcs33~(r.m.s.)$.  We find that 130 X-ray sources have an IR 
counterpart within  $1''~(3\sigma)$.
The source positions, counts and infrared counterparts 
are given in Table 1.
  Thus about 85 \% of the X-ray sources have an IR  
counterpart.  Inversely, 260 or about 2/3 of the IR sources in the
catalogue (Carpenter et al. 1997) emit no 
significant X-rays.  

\subsection{{\itshape K} vs.\ {\itshape H$-$K} magnitude relation} 

Figure 2 shows the $K$ vs.\ $H-K$ magnitude relation for
the X-ray detected (open circles)
and non-detected (filled circles) sources using the IR  data of Carpenter et al. 
(1997).  In general, the X-ray detected sources have more luminous {\itshape
K}-band flux  than those of non-X-ray sources. 
To estimate approximate stellar mass, we show a model track of 2.5~M$_\odot$ stars 
in the age from  0.07~Myr to 2~Myr (solid line; D'Antona \& Mazzitelli 1994).
The extinction effect for 2.5~M$_\odot$ stars with ages of 0.07~Myr and 2~Myr 
are given by dashed lines.
 From Figure 2, we  find that 6 IR stars  are well above the 2.5 M$_\odot$ line
of any ages,  hence these would  be the highest mass stars in the cloud.  
Since IRS~1SW has been identified to be a B0
star (15~M$_\odot$) in ZAMS (Aspin \& Walther 1990), the other 5 sources would
have nearly  equal or even higher mass than 15~M$_\odot$.  
Among the six high mass YSOs, four (IRS~1SW, IRS~2, IRS~3NE and ${\rm a_S}$) 
are found to emit X-rays.     

\subsection{Time  Variability and X-ray Spectra} 

 We make light curves  for all the X-ray sources, then the time 
variability is examined with the Kolmogorov-Smirnov test (Press et al.\ 1992)
 for constant 
flux hypotheses using {\it lcstats} in the {\bfseries XRONOS} (Ver 5.16)
 package\footnote{see http://xronos.gsfc.nasa.gov}.
The significance level of the time variability is listed in Table
1. About a half of the
sources exhibit time variability with 90 \% confidence level.
As for the four high mass stars, three show  time variability with the 98 \%
confidence level.  

For brighter sources with more than 20 counts, we fit the spectrum 
with a thin thermal plasma model (Mewe, Gronenschild, \& van den Oord
1985) using {\bf XSPEC} (Ver. 11.0)\footnote{see http://xspec.gsfc.nasa.gov}.
Since statistics are still limited, we  fix the abundances to be  0.3 solar,
according to the previous reports  (Kamata et al 1997., Yamauchi et
al. 1996). The best-fit parameters are listed in Table 1.
This simple model is acceptable for most of the sources.
The errors are 90 \% confidence limit for the relevant one parameter
(in the range of the $\chi^2$ minimum + 2.7).
The mean temperature is 3~keV and about a half of 
the best-fit temperatures fall between 2~keV and  5~keV.  
The  absorption column density scatters more largely 
from $10^{21}$ to   $10^{23} {\rm cm^{-2}}$. 

For the fainter sources, we fit the spectrum with a fixed-temperature of 3~keV
(the averaged temperature of the brighter sources),  and the best-fit
luminosity and column density are listed in Table 1. 
These values change typically $\pm$50\% and $\pm$30\%,
allowing the temperature to 2~keV and 5~keV, respectively.

\subsection{Individual High Mass Sources} 

 To study  the X-ray properties of four high mass stars 
(IRS~1SW, IRS~2, IRS~3NE and ${\rm a_S}$)
in further detail, we show a closed-up  version of the X-ray images of an area 
$50'' \times 50''$ (Figure 3).   
The X-ray spectrum of IRS~1SW (No.\ 79 in Table 1)
is  fitted with a thin thermal plasma model as is shown in Figure 4a.
The best-fit plasma temperature, absorption column density and X-ray luminosity
are  $\sim$ 2~keV,  $\sim5\times10^{22}$~cm$^{-2}$, and  $10^{31}$ ergs~s$^{-1}$, 
respectively.  The X-rays are time variable with a flare-like event 
as is  shown in Figure 5a.

 The X-ray spectrum and light curve of another candidate of the high mass ZAMS  
star ${\rm a_S}$ (No.\ 91) are given in Figure 4b and 5b, respectively.
The X-ray spectrum is well fitted with a thin thermal plasma model
(see Figure 4b) with the best-fit plasma temperature, absorption column 
and X-ray luminosity of  $\sim$ 2~keV, $\sim5\times10^{22}$~cm$^{-2}$, 
and $6\times 10^{30}$ ergs~s$^{-1}$, respectively. A  flare-like event
is also  observed (see Figure 5b).   
These X-ray features  are very similar to those of IRS~1SW.

The X-ray spectrum and light curve of IRS~2 (No.\ 67) 
are shown in Figure 4c and 5c, respectively. 
The spectrum is fitted with a thin thermal  plasma of 10.9~keV temperature and  absorption 
column of $\sim 10^{23}~{\rm cm}^{-2}$ (see Figure 4c).
The X-ray flux is  highly variable with a slow-rise profile at
the peak luminosity of $2\times 10^{31} {\rm ergs~s^{-1}}$ (see Figure 5c).

 In contrast to the other high mass YSO candidates, the spectrum of IRS~3NE (No.\ 116)
shows no large absorption with the best-fit  $N_{\rm H}$ of 
$1.4\times 10^{22}~{\rm cm^{-2}}$, which is inconsistent with the $H-K$ values
(see the next section).
 One possibility is that the spectrum is contaminated by other unknown sources. 
In  Figure 3, we find a faint diffuse or multiple sources near  IRS~3NE.
We thus re-fit the spectrum with a
strongly absorbed high temperature component plus a less absorbed
low temperature component; the former is likely from IRS~3NE, and the latter would 
be another source around. 
 For the spectral fitting, we  assume the $N_{\rm H}$ of IRS~3NE  to be $7 \times
10^{22}~{\rm cm^{-2}}$ using the relation of Figure 6 (see the next section).
The best-fit two-temperature model is given in Figure 4d.
The high temperature component (that from IRS~3NE) has $kT$ $>$ 
3.3~keV and the luminosity is estimated to be  $1.6\times 10^{30}~{\rm
ergs~s^{-1}}$.   Note that the values in Table 1 are the results of one-component fit, hence are 
different from those of  the two-component fit.  
Since the soft X-ray band of IRS~3NE is largely contaminated by nearby sources, 
we make the X-ray light curve in the hard (2--10~keV) band and is  given in Figure
5d.  Unlike the other high mass YSOs, the light curve shows no significant time 
variability.  However,  the poor statistics can not give strong constraint on the 
time variability.

\section{DISCUSSION} 

\subsection{Global Features}  

 In Figure 6, we plot the best-fit $N_{\rm H}$  as a
function of $H-K$ magnitude, where the bright sources (temperatures are
free parameters)
are given  by circles and those of faint (fixed  temperature of 3~keV) are
shown by boxes. 
The best-fit relation for the brighter sources with more than 60 counts
(filled circles) 
is $N_{\rm H} = [(2.04\pm 0.26) (H-K) - (0.64 \pm 0.26)]\times 10^{22}~{\rm cm^{-2}}$.

Although the intrinsic $H-K$ value for the stars change with their
ages and masses, we assume their intrinsic colors are the same.
Then, from  the relation between $A_V$ and infrared extinction
(Cohen et al.\ 1981), we obtain a relation of $N_{\rm H}
= (1.32 \pm 0.17)\times
10^{21} A_V ~{\rm cm^{-2}}$, significantly different from  that in our Galaxy, 
$N_{\rm H} = 1.79 \times 10^{21} A_V~{\rm cm^{-2}} $ (Predehl \& Schmitt 1995).
This means that the dust-to-gas ratio in this cloud is larger than the mean 
interstellar medium. 
We also derived the average intrinsic color to be $H_0-K_0\sim0.3$.

We estimate extinction corrected
$K$-magnitude ($K$c) using the standard relation of $A_K = 1.38E(H-K)$
(Cohen et al.\ 1981). 
Figure 7 is a scatter plot of $L_{\rm X}$ vs.\ $K$c, where the symbols are  
the same as in Figure 6. 
Although the conversion of $K$c to the stellar mass is not unique, the smaller
$K$c represents, as the first approximation,  the higher mass stars. 
From Figure 7, we find that $L_{\rm X}$  seems to saturate at  $\sim
10^{31}$~ergs~s$^{-1}$ in the high mass end.

\subsection{High Mass YSOs} 

We  discover heavily absorbed X-rays from high mass YSOs (ZAMS and PMS),
IRS~1SW, IRS~2, IRS~3NE and ${\rm a_S}$, but no X-rays from IRS~3SW and IRS~5.
Garmire et al. (2000) reported the detection of X-rays from ``Source~n'' in 
the Orion Nebula. The IR luminosity is comparable to 
our predicted value of high mass stars in  Mon R2. Since ``Source~n'' is known to have
an \ion{H}{2} region (Gezari, Backman, \& Werner 1998), it would be already a
MS star, possibly ZAMS. The
other X-ray emitting high mass stars (Schulz et al. 2001), the Orion Trapezium, 
are also well in the MS phase.
Therefore our observation of Mon R2  provides the first
reliable detection of X-rays from high mass YSOs in the cloud cores.  

IRS~1SW and ${\rm a_S}$ are  strong and heavily 
absorbed X-ray sources, consistent with their
$H-K$ values (see Figure 2). 
 The  high temperature plasma of $\sim$ 2~keV, large absorption
and rapid time variability, in particular flares, are  typical of the embedded 
low mass stars, which show magnetic activity (e.g. Imanishi et al.\ 2001). 
In fact, X-rays from high mass MS stars,
originated in the stellar wind activity, exhibit lower temperature plasma 
of $\sim$1~keV and  relatively  stable light curve (Bergh\"ofer et al.\ 1997). 
 The   X-ray luminosities  are  comparable
to the empirical relation of $L_{\rm X} \sim 10^{-6}$--$10^{-7}\times L_{\rm bol}$
found for stellar wind origin of high mass MS stars 
(Bergh\"ofer et al.\ 1997), but are 
significantly larger  than those of low mass YSOs (Feigelson et al.\
1993; Casanova et al.\ 1995).
 We thus suspect IRS~1SW and ${\rm a_S}$ possibly in ZAMS (Aspin \&
 Walther 1990), 
still dominate the magnetic activity over that of the stellar wind found 
in high mass MS stars.

 IRS~2 exhibits the highest plasma temperature and the largest absorption 
column density among the bright sources in the Mon R2 cloud.
These large values have been only found in the class I low mass stars
(Koyama et al. 1996; Imanishi et al. 2001).
 The relation between the X-ray luminosity and the bolometric luminosity
 of IRS~2 is  also comparable
to the empirical relation for stellar wind origin of high mass MS stars 
(Bergh\"ofer et al.\ 1997).
  Although possible source confusion can not be excluded, 
{\itshape ASCA}  found a big flare from the position of IRS~2, with 
the  peak luminosity of  $10^{33}~{\rm ergs~s^{-1}}$ (Hamaguchi et al.\ 2000). 
 Since the time scale of X-ray variations from IRS~2 (Figure 5c) is long enough to be 
caused by rotational 
modulation,  it may be conceivable 
that X-rays arise from the interaction  of stellar wind with
magnetosphere (e.g.\ Gagn\'e et al.\ 1997).
However,  IRS~2  has no \ion{H}{2} region, hence no strong UV field, 
nor strong stellar wind.  Therefore, the more likely scenario
is that  X-rays from IRS~2 are  due to a solar-like magnetic reconnection. 

We find heavily-absorbed hard X-rays from IRS~3NE in the outflow source IRS~3 
complex (NE and SW). The X-ray luminosity has a relation of $L_{\rm X} \sim
10^{-8} \times L_{\rm bol}$, significantly smaller than the other stars.
Since there are no reliable {\itshape K}-band photometry data of IRS~3NE and SW at 
present, we estimate their {\itshape K}-band magnitude from the 2MASS
data\footnote{http://www.ipac.caltech.edu/2mass/releases/second/doc/explsup.html}, 
assuming they have the same {\itshape K}-band magnitude.
Then IR extinction  is very large in the range of $H-K = 3-4$~mag.  
The outflows found from  IRS~3  would be  derived by 
magnetic fields,  and the infrared polarimetry revealed  that the 
interstellar magnetic field is compressed from  the neighbor of the GMCs
(e.g. Yao et al.\ 1997),  
therefore the X-rays  due to the magnetic reconnection 
between  the accretion disk and the star
surface are conceivable (Hayashi, Shibata, \& Matsumoto 1996).  
To confirm this scenario, a flare detection is essential. Although we 
find a hint of rapid flares in Figure 5d,  these are not statistically 
significant. 

We have suggested  that the high mass  PMS have magnetically driven activity 
similar to that seen in low mass PMS  stars. 
Much of this interpretation arises from the high plasma temperatures and 
flare-like behavior of the X-ray time series (Figure 5).
Since the majority of stars in the universe
seem to form binary pairs, one may argue that these high mass YSOs are  
binaries with low mass companions, and
the low mass YSOs  may be the origin of the X-ray flares and, in some cases, 
the entire X-ray flux.  
 The X-ray luminosity of these high mass YSOs, $\sim
10^{30-31}$~ergs~s$^{-1}$, are however 
significantly larger than those of typical low mass YSOs of $\sim
10^{28-29}$~ergs~s$^{-1}$.  Thus contribution of low mass companion, if any,
may be small fraction of the bulk X-rays observed from the high mass  YSOs. 

 The high mass YSOs in Mon R2 can be compared to the Orion Trapezium stars,
$\theta^1$ Ori A--E.
Since the Trapezium stars are producing luminous \ion{H}{2} regions, they already 
entered the MS phase with very strong stellar winds.
Schulz et al.\ (2001)
reported that three of the trapezium stars,  C (O7V), E (B0.5) and A (B0V-B1V) exhibit
X-rays with the luminosities of about $2\times 10^{32}$~ergs~s$^{-1}$, 
$4\times 10^{31}$~ergs~s$^{-1}$, and $2\times 10^{31}$~ergs~s$^{-1}$, respectively. 
They also  show a high temperature component of about 2~keV in addition
to a lower temperature component of about 1~keV. 
 The luminosities  and temperatures of the soft components 
are consistent with the stellar wind origin. For the hard components, however,
the plasma  temperature is higher than that predicted with  
the stellar wind velocity, 
hence would be  attributable to a magnetic origin, like the Mon R2 stars.
Moreover, Gagn\'e et al. (1997) found X-ray variations from $\theta^1$ Ori C. 
 This would be  supporting evidence for remaining the magnetic activity on
these high mass stars, although Gagn\'e et al. (1997)
interpreted it as the interaction
of a magnetosphere and the stellar winds. 

 Thus we propose a working hypothesis  for the study of X-ray evolution of young stars 
which was proposed in the X-ray study of intermediate mass young stars
Herbig Ae/Be stars (Hamaguchi 2001), 
that, even high mass stars, produce variable and hard (2--3~keV) X-rays due to the magnetic
activity between the accretion disk and the stellar surface 
in a PMS phase (IRS~2 and IRS~3NE). It continues until a ZAMS phase 
(IRS~1SW and ${\rm a_S}$), then gradually 
stellar wind activities come in (the Trapezium stars). The
magnetic activity fades as the accretion disk disappears. Finally most
of high mass MS stars emit soft (1~keV) X-rays by the strong stellar wind.

\begin{figure}[thbp]
\plotone{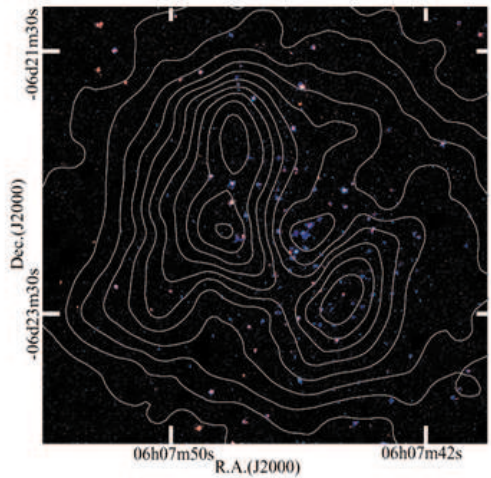}
\caption{ACIS-I image of the Mon R2 cloud. Red and blue colors
represent photons in the soft (0.5--2.0 keV) and hard (2.0--10.0 keV)
X-ray bands, respectively. CS $J = 5 \rightarrow 4$ line intensity map
 (Choi et al.\ 2000) is overlaid (white contours).
}
\end{figure}

\begin{figure}
\plotone{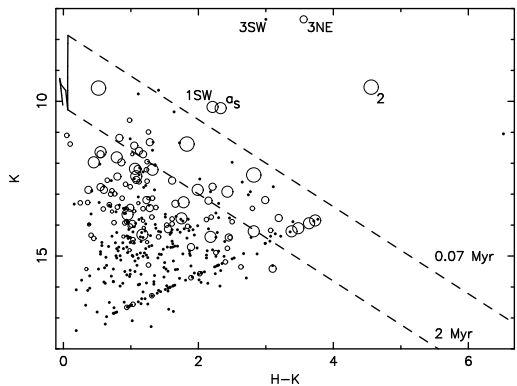}
\caption{Plot of the {\itshape K}-mag and $H-K$ mag relation. For the {\itshape H}-band
non-detection sources, we assume their $H$-band magnitude to be  17.6, the completeness
limit (Carpenter et al.\ 1997). Open circles are the X-ray detected, 
and filled circles are non-detected sources. The radius of the open circles
gives logarithmic of the X-ray flux.
The solid and dashed lines represent time history of stars of 2.5~M$_\odot$
and reddening vectors at the ages of 0.07 and 2~Myr, respectively.}
\end{figure}

\begin{figure}
\plotone{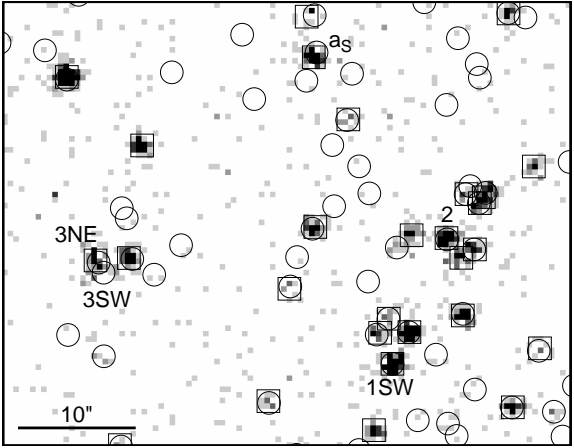}
\caption{Expanded view of the cloud core region ($50'' \times 50''$). Circles and squares are
plot of the infrared and X-ray source positions, respectively. 
The four X-ray bright high mass YSOs (see text) are indicated by 
${\rm a_S}$, 1SW, 2, 3NE and 3SW.}
\end{figure}

\begin{figure}
\plotone{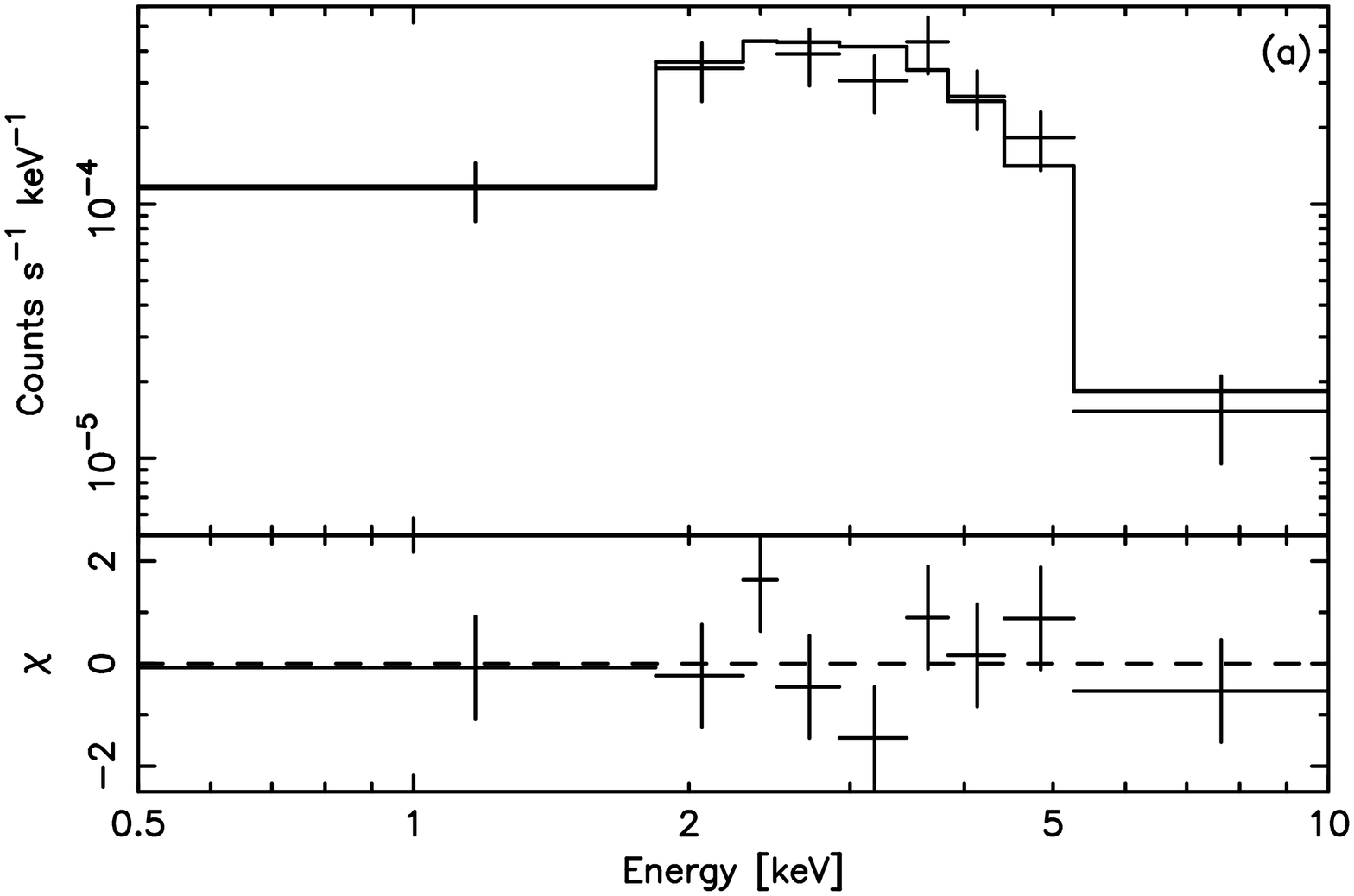}
\plotone{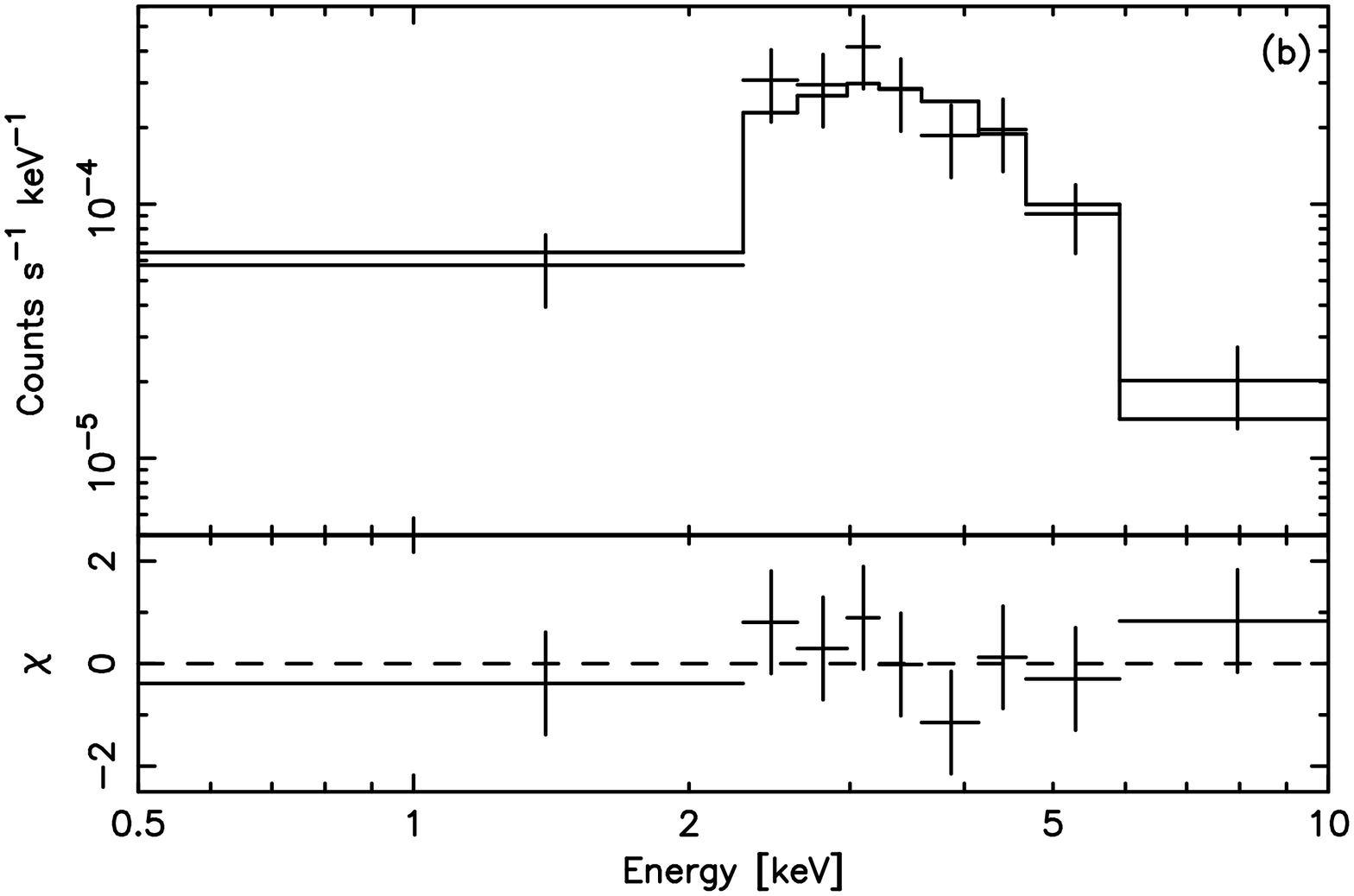}
\end{figure}
\begin{figure}
\plotone{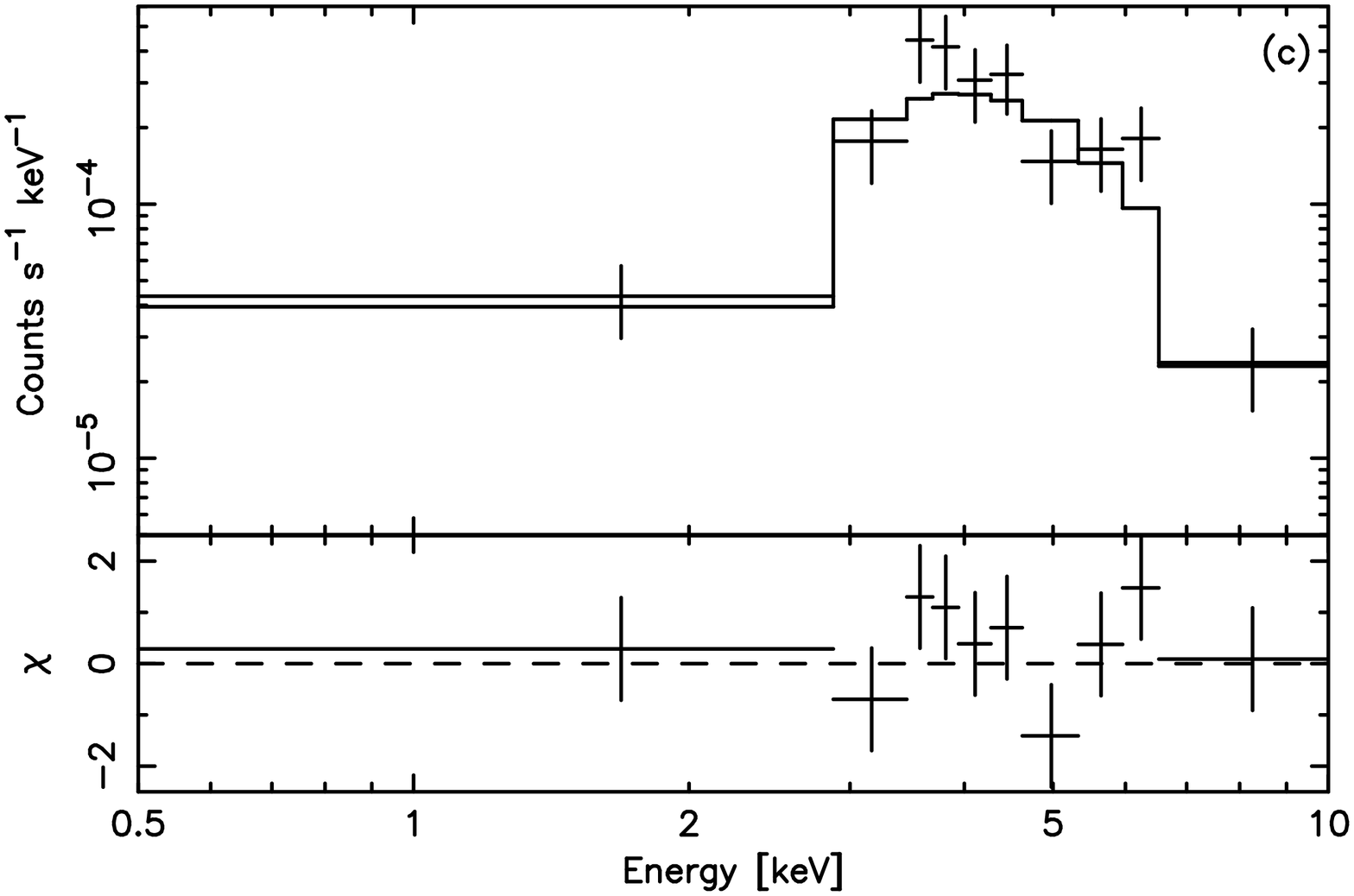}
\plotone{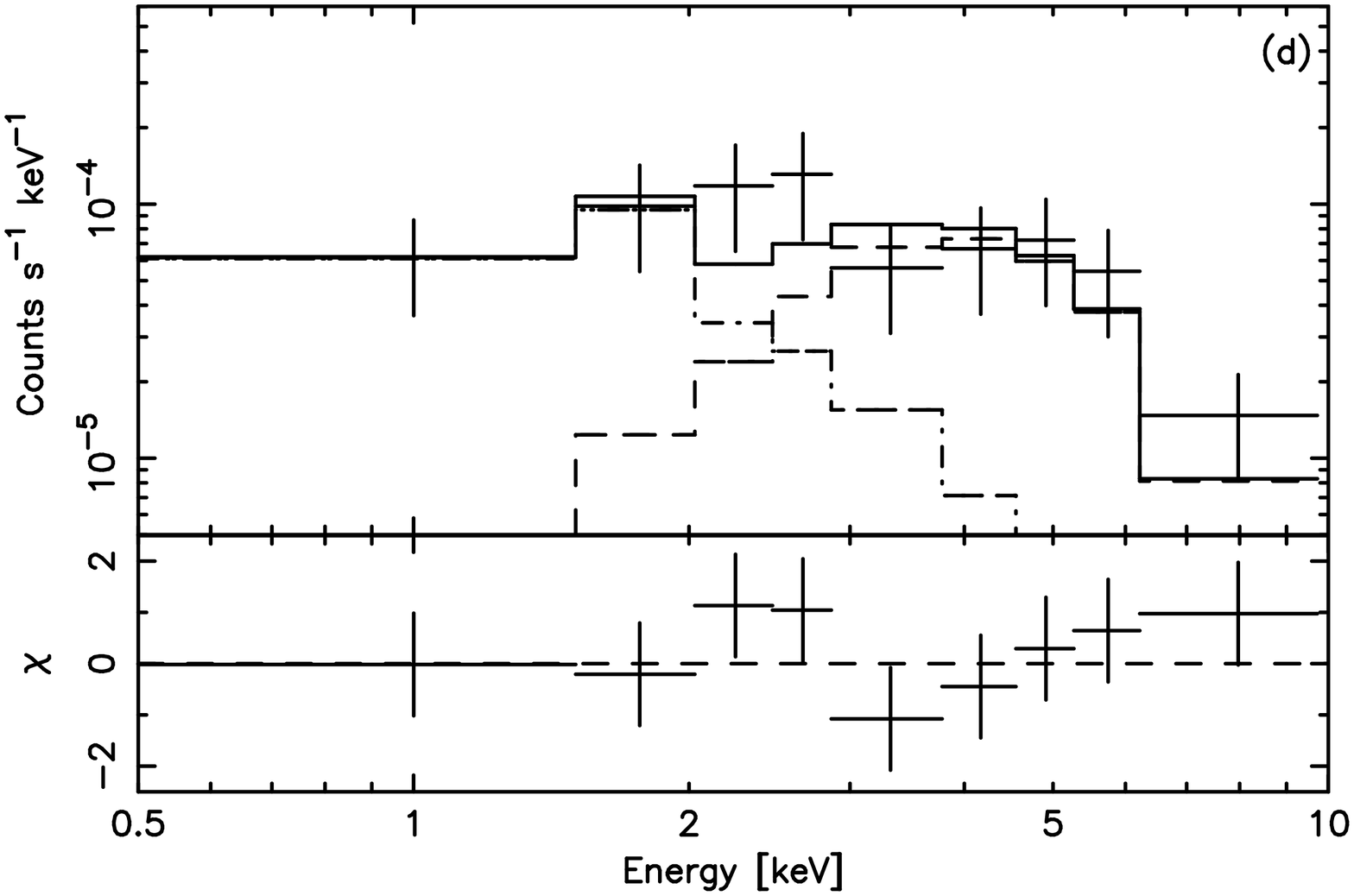}
 \caption{X-ray spectra and the best-fit thin thermal model for  high mass stars.
The best-fit two component model is shown for IRS~3NE.
(a):~IRS~1SW; (b):~${\rm a_S}$; (c):~IRS~2; (d):~IRS~3NE}
\end{figure}

\begin{figure}
\plotone{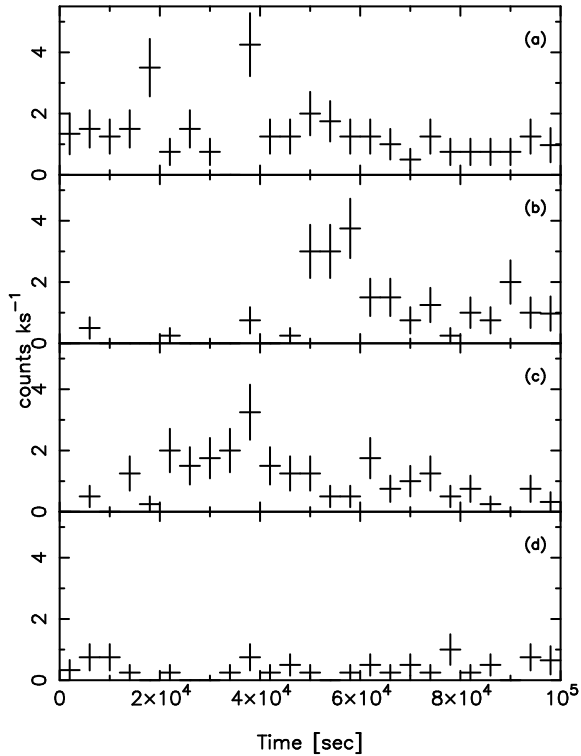}
\caption{Light curves of the high mass YSOs in
the 0.5--10~keV  band. For IRS~3NE (d), the energy band is 2--10~keV. 
 The horizontal axis is the time from
the observation start (MJD = 51880.97) and the vertical axis is the
ACIS-I counts per ks.   (a):IRS~1SW; (b):${\rm a_S}$; (c):IRS~2; (d):IRS~3NE
}
\end{figure}

\begin{figure}
\plotone{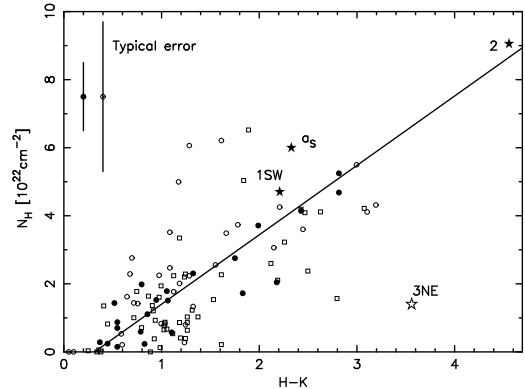}
\caption{Plots of the absorption column ($N_{\rm H}$) and the
near-infrared color ($H-K$) relation.  Stars, circles and boxes represent 
the bright sources (more than 20~counts, temperature is free parameters) and 
faint sources (less than 20~counts, temperature is fixed to 3~keV, see text), 
respectively. The solid line is the 
best-fit relation using the bright sources (filled circles) 
with more than  60~counts.
Typical 1 $\sigma$ errors of $N_{\rm H}$ are separately shown for the filled
circles (more than 60~counts) and open circle (20--60~counts) sources.}

\end{figure}

\begin{figure}
\plotone{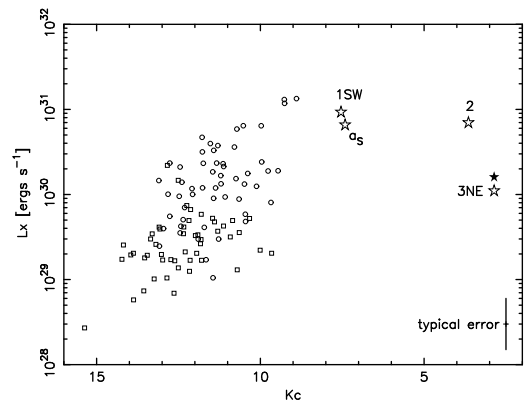}
\caption{Plots of the extinction corrected $K$-mag ($K$c) and $L_{\rm X}$
relation. Symbols are the same as in Figure 6. The filled star 
represent the result of IRS~3NE with a 2-components model (see text).
Typical 1 $\sigma$ errors of $L_{\rm X}$ in logarithmic scale
is also shown.}
\end{figure}

\clearpage
\tabletypesize{\scriptsize}
\catcode`?=\active \def?{\phn}
\begin{deluxetable}{rllrrlclrcrc}
\tablewidth{533pt}
\tablenum{1}
\tablecaption{{\itshape Chandra} X-ray sources in the Monoceros R2 region \label{tab:target}}
\tablehead{
\colhead{No.} & \colhead{R.A.$^{\rm a}$} & \colhead{Dec.$^{\rm a}$} & 
 \colhead{Cts}& \colhead{V.P.$^{\rm b}$}& \colhead{$kT$} & \colhead{$\log({\rm EM})$} &
 \colhead{$N_{\rm H}$} & \colhead{Flux$^{\rm c}$} &
 \colhead{$\log(L_{\rm X})^{\rm d}$} &
\colhead{IR$^{\rm e}$} & \colhead{Names$^{\rm f}$}\\
\colhead{} & \colhead{(2000)} & \colhead{(2000)} & \colhead{} &
\colhead{(\%)} & \colhead{(keV)} & \colhead{(${\rm cm^{-3}}$)} &
 \colhead{($\times 10^{22}~{\rm cm^{-2}}$)} & \colhead{} &
 \colhead{(${\rm ergs~s^{-1}}$)} &
\colhead{} & \colhead{}
}
\startdata
 1	&	06 07 40.53 &	-06 22 45.7 &	16 &	44.0 &	3.0(fixed) &	53.5(53.1--53.9) &	1.6(0.6--3.4) &	2.0 &	29.5(29.2--29.7) &	1&\\
 2	&	06 07 40.57 &	-06 23 56.7 &	26 &	99.2 &	3.1($>$0.9) &	53.7(53.2--54.7) &	2.0(0.7--4.6) &	3.5 &	29.8(29.6--30.5) &	&\\
 3	&	06 07 40.63 &	-06 22 59.2 &	14 &	100.0 &	3.0(fixed) &	53.7(53.2--54.2) &	3.9(1.7--7.9) &	2.2 &	29.7(29.3--29.8) &	&\\
 4	&	06 07 40.85 &	-06 24 04.8 &	22 &	84.4 &	1.6(0.7--9.9) &	54.0(53.3--54.8) &	2.2(0.9--4.5) &	2.3 &	29.9(29.5--30.7) &	6&\\
 5	&	06 07 40.87 &	-06 21 17.2 &	22 &	60.2 &	\llap{1}2.8($>$1.2) &	53.3(53.1--54.2) &	0.8(0.2--3.0) &	3.7 &	29.6(29.4--30.0) &	3&\\
 6	&	06 07 40.96 &	-06 24 09.9 &	11 &	83.3 &	3.0(fixed) &	53.2(52.7--53.6) &	1.0(0.1--3.6) &	1.3 &	29.3(28.8--29.6) &	9&\\
 7	&	06 07 41.07 &	-06 24 08.8 &	30 &	89.7 &	2.6($>$0.5) &	54.1(53.2--56.5) &	4.3(0.5--73.2) &	4.4 &	30.1(29.6--37.0) &	13&\\
 8	&	06 07 41.38 &	-06 22 12.4 &	99 &	99.5 &	1.6(1.2--2.5) &	54.9(54.5--55.2) &	3.7(2.6--4.8) &	14.2 &	30.8(30.6--31.1) &	19&\\
 9	&	06 07 42.10 &	-06 23 24.5 &	9 &	30.4 &	3.0(fixed) &	53.6(53.0--54.3) &	5.3(1.4--16.6) &	1.8 &	29.7(29.1--30.0) &	&\\
 10	&	06 07 42.13 &	-06 22 55.3 &	24 &	84.7 &	4.8($>$1.1) &	53.6(53.2--54.5) &	1.8(0.8--4.5) &	3.9 &	29.8(29.6--30.3) &	25&\\
 11	&	06 07 42.20 &	-06 23 10.9 &	10 &	57.1 &	3.0(fixed) &	53.6(53.0--54.2) &	4.1(1.3--22.4) &	1.9 &	29.7(29.1--30.2) &	27&\\
 12	&	06 07 42.22 &	-06 23 20.3 &	10 &	63.3 &	3.0(fixed) &	54.7(54.0--55.3) &	\llap{4}7.6(17.1--105.9) &	5.2 &	30.8(30.1--31.3) &	&\\
 13	&	06 07 42.27 &	-06 22 53.6 &	25 &	78.0 &	4.2($>$1.4) &	53.8(53.4--54.5) &	3.6(1.3--7.9) &	4.6 &	30.0(29.7--30.3) &	29&\\
 14	&	06 07 42.40 &	-06 23 49.8 &	19 &	58.7 &	3.0(fixed) &	53.2(52.5--53.5) &	1.4(0.1--3.0) &	1.2 &	29.3(28.6--29.4) &	32&\\
 15	&	06 07 42.43 &	-06 23 14.6 &	7 &	93.7 &	3.0(fixed) &	53.5(52.0--53.9) &	5.0(0.0--20.5) &	1.2 &	29.5(28.3--29.8) &	34&\\
 16	&	06 07 42.73 &	-06 23 07.7 &	14 &	99.5 &	3.0(fixed) &	53.2(52.8--53.6) &	1.0(0.2--3.3) &	1.3 &	29.3(28.9--29.6) &	39&\\
 17	&	06 07 42.84 &	-06 23 11.5 &	17 &	89.3 &	3.0(fixed) &	53.6(53.2--53.9) &	2.8(1.2--6.9) &	2.3 &	29.7(29.3--29.9) &	43&\\
 18	&	06 07 42.96 &	-06 22 48.3 &	9 &	100.0 &	3.0(fixed) &	54.3(53.3--55.6) &	\llap{2}5.6(4.7--160.5) &	3.1 &	30.3(29.4--31.7) &	47&\\
 19	&	06 07 42.97 &	-06 23 21.1 &	13 &	32.6 &	3.0(fixed) &	53.4(52.9--53.8) &	2.2(0.6--8.6) &	1.6 &	29.5(29.0--29.8) &	48&\\
 20	&	06 07 42.99 &	-06 22 33.8 &	13 &	36.3 &	3.0(fixed) &	53.2(52.5--53.8) &	0.9(0.0--7.4) &	1.2 &	29.2(28.7--29.8) &	50&\\
 21	&	06 07 43.00 &	-06 24 10.4 &	27 &	100.0 &	3.8($>$0.7) &	54.1(53.5--56.8) &	6.2(1.9--39.6) &	6.1 &	30.2(29.8--32.7) &	54&\\
 22	&	06 07 43.10 &	-06 22 06.4 &	19 &	96.7 &	3.0(fixed) &	53.7(53.1--53.9) &	2.2(0.6--4.4) &	2.7 &	29.7(29.2--29.9) &	56&\\
 23	&	06 07 43.11 &	-06 23 56.9 &	13 &	93.9 &	3.0(fixed) &	53.2(52.7--53.5) &	0.9(0.1--3.4) &	1.2 &	29.3(28.8--29.6) &	58&\\
 24	&	06 07 43.23 &	-06 22 58.0 &	13 &	50.5 &	3.0(fixed) &	53.9(53.5--54.2) &	7.0(3.2--15.2) &	2.9 &	30.0(29.6--30.2) &	&\\
 25	&	06 07 43.54 &	-06 22 44.7 &	16 &	38.9 &	3.0(fixed) &	53.2(52.8--53.5) &	1.0(0.3--2.2) &	1.3 &	29.3(28.9--29.4) &	66&\\
 26	&	06 07 43.58 &	-06 22 51.7 &	70 &	99.5 &	3.1(1.6--6.5) &	54.1(53.8--54.4) &	1.1(0.7--1.8) &	8.7 &	30.1(30.0--30.3) &	70&${\rm d_N}$\\
 27	&	06 07 43.58 &	-06 23 21.8 &	30 &	98.2 &	?\phd?($>$2.2)$^{\rm g}$ &	53.9(53.7--54.2) &	5.6(3.0--12.5) &	10.0 &	30.2(30.0--30.3) &	65&\\
 28	&	06 07 43.61 &	-06 23 10.6 &	14 &	84.5 &	3.0(fixed) &	53.5(53.1--53.8) &	1.9(0.9--4.1) &	2.1 &	29.6(29.3--29.8) &	75&\\
 29	&	06 07 43.66 &	-06 24 21.7 &	10 &	87.1 &	3.0(fixed) &	53.2(52.5--53.6) &	1.0(0.1--4.4) &	1.1 &	29.2(28.6--29.6) &	78&\\
 30	&	06 07 43.69 &	-06 23 01.6 &	30 &	61.8 &	6.9($>$1.3) &	54.3(53.9--55.9) &	\llap{2}0.4(7.7--53.9) &	11.0 &	30.6(30.1--31.5) &	&\\
 31	&	06 07 43.72 &	-06 22 46.8 &	26 &	53.3 &	1.4(0.8--2.6) &	54.1(53.5--54.8) &	2.3(1.0--4.5) &	2.6 &	30.0(29.6--30.6) &	79&\\
 32	&	06 07 43.86 &	-06 23 27.1 &	9 &	99.2 &	3.0(fixed) &	53.2(52.5--53.9) &	1.6(0.1--12.9) &	1.2 &	29.3(28.6--29.9) &	86&\\
 33	&	06 07 43.92 &	-06 23 13.3 &	36 &	94.8 &	3.6($>$1.2) &	54.2(53.7--55.2) &	5.5(2.2--12.3) &	7.7 &	30.3(30.0--31.0) &	87&\\
 34	&	06 07 43.96 &	-06 23 27.1 &	10 &	56.0 &	3.0(fixed) &	53.7(53.0--54.3) &	5.4(1.5--28.1) &	2.0 &	29.7(29.1--30.3) &	&\\
 35	&	06 07 44.14 &	-06 24 20.5 &	44 &	63.0 &	1.3(0.8--2.2) &	54.7(54.1--55.3) &	4.1(2.8--6.4) &	5.6 &	30.6(30.2--31.2) &	93&\\
 36	&	06 07 44.21 &	-06 23 32.0 &	14 &	99.4 &	3.0(fixed) &	54.1(53.7--54.4) &	\llap{1}1.9(5.5--23.8) &	3.6 &	30.2(29.8--30.4) &	&\\
 37	&	06 07 44.21 &	-06 23 25.4 &	14 &	65.6 &	3.0(fixed) &	53.4(52.9--53.7) &	1.2(0.3--3.4) &	1.7 &	29.4(29.0--29.7) &	95&\\
 38	&	06 07 44.23 &	-06 22 30.7 &	33 &	95.7 &	\llap{2}2.9($>$1.0) &	53.3(53.0--53.5) &	0.2(0.0--2.0) &	3.8 &	29.5(29.5--30.1) &	94&\\
 39	&	06 07 44.29 &	-06 24 22.5 &	7 &	89.1 &	3.0(fixed) &	52.8(52.0--53.1) &	0.5(0.0--2.3) &	0.6 &	28.9(28.5--29.0) &	97&\\
 40	&	06 07 44.46 &	-06 22 41.3 &	42 &	96.6 &	2.7(1.2--17.8) &	54.0(53.6--54.7) &	2.2(0.9--4.3) &	5.7 &	30.1(29.9--30.6) &	103&\\
 41	&	06 07 44.49 &	-06 22 33.8 &	273 &	100.0 &	4.3(2.8--16.7) &	55.1(54.8--55.3) &	5.9(4.4--7.1) &	72.7 &	31.2(31.1--31.3) &	&\\
 42	&	06 07 44.52 &	-06 23 21.7 &	24 &	66.1 &	2.1($>$0.4) &	53.6(53.0--55.7) &	1.3(0.2--4.9) &	2.0 &	29.6(29.4--31.5) &	105&\\
 43	&	06 07 44.53 &	-06 22 53.0 &	12 &	92.3 &	3.0(fixed) &	53.9(53.4--54.3) &	7.2(2.9--22.6) &	2.8 &	30.0(29.5--30.3) &	&\\
 44	&	06 07 44.60 &	-06 23 15.5 &	14 &	58.6 &	3.0(fixed) &	53.7(53.2--54.0) &	4.2(1.9--9.6) &	2.3 &	29.8(29.4--29.9) &	107&\\
 45	&	06 07 44.73 &	-06 23 42.3 &	92 &	99.1 &	1.9(1.1--2.8) &	54.6(54.2--55.0) &	2.3(1.7--3.6) &	11.5 &	30.5(30.4--30.9) &	&\\
 46	&	06 07 44.79 &	-06 24 05.2 &	12 &	66.8 &	3.0(fixed) &	53.3(52.7--53.6) &	1.6(0.4--4.2) &	1.3 &	29.3(28.8--29.5) &	118&\\
 47	&	06 07 44.79 &	-06 23 26.1 &	5 &	84.7 &	3.0(fixed) &	53.4(51.9--54.0) &	4.4(0.0--35.7) &	1.1 &	29.4(28.3--30.0) &	122&\\
 48	&	06 07 44.81 &	-06 23 35.9 &	16 &	63.5 &	3.0(fixed) &	53.9(53.5--54.2) &	6.5(3.4--13.4) &	3.3 &	30.0(29.7--30.2) &	116&\\
 49	&	06 07 44.94 &	-06 22 40.0 &	21 &	83.6 &	3.2(0.9--19.2) &	53.4(53.0--54.1) &	0.8(0.1--2.5) &	2.2 &	29.5(29.3--29.9) &	119&\\
 50	&	06 07 44.95 &	-06 23 26.0 &	5 &	62.3 &	3.0(fixed) &	53.1(51.9--53.6) &	2.3(0.0--17.5) &	0.8 &	29.2(28.3--29.6) &	153&\\
 51	&	06 07 45.22 &	-06 23 28.3 &	54 &	80.3 &	2.7(1.3--9.0) &	54.2(53.8--54.7) &	2.5(1.5--4.0) &	8.5 &	30.3(30.1--30.6) &	131&\\
 52	&	06 07 45.23 &	-06 23 03.3 &	12 &	49.0 &	3.0(fixed) &	53.7(53.1--54.6) &	4.2(1.4--45.0) &	2.1 &	29.7(29.2--30.7) &	132&\\
 53	&	06 07 45.25 &	-06 23 19.9 &	10 &	99.2 &	3.0(fixed) &	53.0(52.5--53.3) &	0.6(0.0--1.8) &	1.0 &	29.1(28.6--29.3) &	139&IRS~4\\
 54	&	06 07 45.25 &	-06 23 35.5 &	24 &	97.9 &	3.3($>$0.5) &	53.9(53.3--56.4) &	3.7(1.2--13.9) &	3.9 &	29.9(29.6--32.1) &	134&\\
 55	&	06 07 45.27 &	-06 22 47.2 &	30 &	87.4 &	2.0(0.8--5.2) &	54.3(53.6--55.5) &	4.5(2.4--10.2) &	5.1 &	30.3(29.9--31.4) &	&\\
 56	&	06 07 45.30 &	-06 22 28.3 &	24 &	84.4 &	2.4(0.9--13.8) &	53.3(52.9--54.0) &	0.3(0.0--1.5) &	1.8 &	29.3(29.2--29.9) &	&\\
 57	&	06 07 45.39 &	-06 23 08.5 &	42 &	100.0 &	\llap{5}9.2($>$3.0) &	54.0(53.8--54.2) &	3.7(2.4--7.6) &	12.7 &	30.2(30.1--30.4) &	142&\\
 58	&	06 07 45.42 &	-06 22 33.7 &	33 &	83.9 &	2.4(1.0--8.4) &	54.3(53.7--55.2) &	4.3(2.3--8.8) &	6.2 &	30.3(30.0--31.0) &	143&\\
 59	&	06 07 45.46 &	-06 21 28.4 &	67 &	58.5 &	1.8(0.9--3.0) &	54.3(53.9--55.0) &	1.8(0.9--3.5) &	6.9 &	30.2(30.0--30.9) &	146&\\
 60	&	06 07 45.47 &	-06 23 16.5 &	24 &	84.6 &	1.6(0.7--4.1) &	53.8(53.2--54.5) &	1.4(0.6--3.2) &	2.0 &	29.7(29.4--30.3) &	147&\\
 61	&	06 07 45.57 &	-06 22 49.9 &	86 &	98.5 &	2.0(1.5--3.8) &	54.8(54.4--55.3) &	4.7(3.2--6.8) &	15.5 &	30.8(30.5--31.2) &	150&\\
 62	&	06 07 45.59 &	-06 22 50.5 &	74 &	99.4 &	1.6(0.9--3.5) &	54.8(54.2--55.5) &	4.9(3.2--7.2) &	11.2 &	30.8(30.4--31.4) &	160&\\
 63	&	06 07 45.62 &	-06 22 54.6 &	40 &	100.0 &	?\phd?($>$4.4)$^{\rm g}$ &	54.0(53.8--54.2) &	5.6(3.2--13.4) &	12.3 &	30.3(30.1--30.4) &	167&\\
 64	&	06 07 45.67 &	-06 22 49.7 &	49 &	78.7 &	2.5(1.4--6.9) &	54.4(53.9--55.0) &	4.0(2.4--6.6) &	8.7 &	30.4(30.1--30.9) &	197&${\rm c_S}$\\
 65	&	06 07 45.69 &	-06 23 00.4 &	67 &	99.2 &	2.0(1.1--3.9) &	54.9(54.4--55.6) &	8.0(5.1--12.7) &	14.5 &	30.9(30.5--31.5) &	169&\\
 66	&	06 07 45.70 &	-06 22 55.3 &	50 &	99.8 &	7.8($>$2.0) &	54.2(53.9--54.8) &	5.3(2.9--9.9) &	14.2 &	30.4(30.2--30.7) &	&\\
 67	&	06 07 45.79 &	-06 22 53.6 &	99 &	99.3 &	\llap{1}0.9($>$2.0) &	54.6(54.4--55.3) &	9.1(5.7--16.9) &	35.4 &	30.8(30.7--31.3) &	178&IRS~2\\
 68	&	06 07 45.83 &	-06 21 46.6 &	541 &	98.5 &	2.0(1.7--2.3) &	55.1(55.0--55.3) &	1.4(1.2--1.6) &	60.8 &	31.1(31.0--31.2) &	181&\\
 69	&	06 07 45.88 &	-06 21 54.3 &	34 &	56.3 &	0.5(0.1--0.8) &	53.1(52.9--58.9) &	0.0(0.0--0.9) &	1.2 &	29.0(29.0--31.6) &	184&\\
 70	&	06 07 45.90 &	-06 24 00.9 &	84 &	99.7 &	6.0($>$1.5) &	54.8(54.5--55.5) &	\llap{1}5.0(8.9--27.4) &	33.3 &	31.0(30.8--31.7) &	&\\
 71	&	06 07 45.94 &	-06 23 32.6 &	83 &	74.1 &	2.7(1.6--5.4) &	54.5(54.1--54.8) &	2.8(1.8--4.0) &	13.9 &	30.5(30.3--30.8) &	192&\\
 72	&	06 07 45.99 &	-06 22 53.3 &	34 &	56.0 &	2.5($>$0.6) &	54.3(53.7--56.1) &	6.2(2.5--13.2) &	6.7 &	30.4(30.0--31.8) &	&\\
 73	&	06 07 46.00 &	-06 22 25.7 &	19 &	77.6 &	3.0(fixed) &	53.4(53.0--53.6) &	0.9(0.4--1.7) &	1.8 &	29.4(29.2--29.5) &	210&\\
 74	&	06 07 46.00 &	-06 23 43.7 &	14 &	76.8 &	3.0(fixed) &	53.2(52.8--53.5) &	0.8(0.2--2.5) &	1.4 &	29.3(28.9--29.5) &	200&\\
 75	&	06 07 46.01 &	-06 23 01.9 &	122 &	100.0 &	2.8(1.6--4.5) &	55.0(54.7--55.5) &	7.3(5.5--10.9) &	30.5 &	31.0(30.8--31.4) &	199&\\
 76	&	06 07 46.01 &	-06 22 24.0 &	40 &	96.6 &	1.5(0.8--2.8) &	54.2(53.7--54.9) &	2.0(0.9--3.7) &	3.9 &	30.1(29.8--30.8) &	195&${\rm c_N}$\\
 77	&	06 07 46.04 &	-06 23 50.2 &	13 &	31.9 &	3.0(fixed) &	53.6(53.2--53.9) &	3.2(1.4--8.1) &	2.2 &	29.7(29.3--29.9) &	203&\\
 78	&	06 07 46.10 &	-06 22 05.5 &	116 &	100.0 &	7.7($>$3.1) &	54.3(54.2--54.6) &	2.3(1.3--3.5) &	25.5 &	30.6(30.5--30.7) &	204&\\
 79	&	06 07 46.11 &	-06 23 04.7 &	130 & 98.8 &	1.9(1.3--2.8) &	55.0(54.7--55.4) &	4.7(3.5--6.4) & 22.2 &	31.0(30.8--31.3) &	207&IRS~1SW\\
 80	&	06 07 46.13 &	-06 23 00.7 &	26 &	53.2 &	2.0($>$0.6) &	54.1(53.4--54.8) &	3.9(0.3--8.8) &	3.3 &	30.1(29.4--31.5) &	237&\\
 81	&	06 07 46.16 &	-06 21 30.7 &	16 &	85.0 &	3.0(fixed) &	53.2(52.8--53.4) &	0.6(0.2--1.4) &	1.5 &	29.3(29.0--29.4) &	211&\\
 82	&	06 07 46.18 &	-06 23 01.9 &	28 & 96.5 &	4.0($>$1.5) &	53.1(52.9--53.3) &	0.0(0.0--0.2) & 2.1 &	29.2(29.2--29.3) &	214& IRS~1NE\\
 83	&	06 07 46.22 &	-06 23 10.6 &	53 &	80.5 &	3.8($>$1.2) &	54.4(53.9--55.6) &	5.8(2.3--18.9) &	12.0 &	30.5(30.2--31.5) &	&\\
 84	&	06 07 46.34 &	-06 23 35.4 &	12 &	61.8 &	3.0(fixed) &	53.3(52.7--53.5) &	1.5(0.3--3.6) &	1.2 &	29.3(28.8--29.5) &	223&\\
 85	&	06 07 46.37 &	-06 22 43.1 &	15 &	93.5 &	3.0(fixed) &	53.2(52.8--53.5) &	0.9(0.3--2.0) &	1.3 &	29.3(28.9--29.4) &	229&\\
 86	&	06 07 46.39 &	-06 22 24.5 &	22 &	41.8 &	?\phd?($>$2.4)$^{\rm g}$ &	53.2(52.9--53.4) &	0.3(0.0--1.0) &	3.2 &	29.5(29.2--29.6) &	228&\\
 87	&	06 07 46.52 &	-06 23 30.2 &	15 &	99.7 &	3.0(fixed) &	53.7(53.3--53.9) &	2.6(1.3--5.3) &	2.6 &	29.7(29.4--29.9) &	290&\\
 88	&	06 07 46.53 &	-06 23 18.5 &	14 &	93.9 &	3.0(fixed) &	54.2(53.4--55.2) &	\llap{1}6.4(4.1--90.8) &	3.3 &	30.2(29.6--31.2) &	&\\
    89	&	06 07 46.54 &	-06 23 26.7 &	20 &	64.7 &	3.0(fixed) &	53.6(53.3--53.8) &	2.4(1.3--4.0) &	2.5 &	29.7(29.4--29.8) &	235&\\
    90	&	06 07 46.56 &	-06 22 52.6 &	45 &	98.5 &	3.0(1.4--21.2) &	54.1(53.6--54.6) &	2.6(1.1--4.3) &	6.8 &	30.1(29.9--30.5) &	241&\\
    91	&	06 07 46.57 &	-06 22 37.6 &	89 &	100.0 &	2.6(1.6--6.2) &	54.8(54.3--55.4) &	6.0(4.0--9.2) &	19.7 &	30.8(30.6--31.2) &	239&${\rm a_S}$\\
    92	&	06 07 46.57 &	-06 22 19.0 &	17 &	99.4 &	3.0(fixed) &	53.8(53.5--54.0) &	3.3(2.0--5.8) &	3.2 &	29.9(29.6--30.0) &	238&\\
    93	&	06 07 46.62 &	-06 22 34.0 &	13 &	100.0 &	3.0(fixed) &	53.5(53.0--53.8) &	2.3(0.9--6.1) &	1.8 &	29.6(29.1--29.8) &	240&${\rm a_N}$\\
    94	&	06 07 46.63 &	-06 23 30.0 &	27 &	90.0 &	2.1(0.7--6.6) &	54.2(53.5--55.7) &	4.3(2.2--11.6) &	4.6 &	30.2(29.9--31.5) &	245&\\
    95	&	06 07 46.68 &	-06 23 12.9 &	74 &	92.5 &	1.9(1.2--3.1) &	54.4(54.1--54.8) &	2.0(1.4--3.0) &	8.9 &	30.4(30.2--30.7) &	248&${\rm b_N}$\\
    96	&	06 07 46.72 &	-06 22 58.0 &	11 &	87.7 &	3.0(fixed) &	53.5(53.0--53.7) &	2.3(0.8--5.0) &	1.7 &	29.5(29.1--29.7) &	249&\\
    97	&	06 07 46.84 &	-06 23 08.0 &	14 &	80.1 &	3.0(fixed) &	53.2(52.7--53.4) &	0.7(0.1--1.8) &	1.3 &	29.2(28.8--29.4) &	250&\\
    98	&	06 07 46.91 &	-06 24 17.3 &	31 &	89.6 &	3.8(1.6--43.4) &	54.0(53.7--54.5) &	1.2(0.6--2.6) &	9.3 &	30.1(30.0--30.4) &	254&\\
    99	&	06 07 47.04 &	-06 21 26.6 &	8 &	74.1 &	3.0(fixed) &	53.1(52.3--53.4) &	0.7(0.0--2.5) &	1.0 &	29.1(28.7--29.3) &	260&\\
   100	&	06 07 47.06 &	-06 23 33.8 &	32 &	100.0 &	2.0(0.4--55.9) &	54.4(53.6--57.3) &	6.1(2.3--17.2) &	5.3 &	30.4(29.9--33.0) &	262&IRS~7\\
   101	&	06 07 47.12 &	-06 21 28.3 &	6 &	67.4 &	3.0(fixed) &	53.0(52.2--53.4) &	0.4(0.0--3.8) &	0.9 &	29.0(28.7--29.4) &	263&\\
   102	&	06 07 47.20 &	-06 22 31.2 &	57 &	84.7 &	5.4(2.3--65.2) &	54.2(53.9--54.6) &	3.5(2.0--5.8) &	13.4 &	30.4(30.2--30.6) &	264&g\\
   103	&	06 07 47.30 &	-06 23 47.2 &	33 &	86.5 &	1.1(0.4--2.1) &	54.1(53.6--55.3) &	1.6(0.8--3.0) &	2.3 &	30.0(29.6--31.2) &	267&\\
   104	&	06 07 47.35 &	-06 21 51.8 &	119 &	64.3 &	2.7(1.6--5.7) &	54.8(54.4--55.2) &	4.1(2.9--5.7) &	23.9 &	30.8(30.6--31.1) &	268&\\
   105	&	06 07 47.48 &	-06 22 18.7 &	26 &	85.8 &	7.2($>$1.0) &	53.7(53.4--55.3) &	3.1(1.2--12.7) &	5.7 &	29.9(29.7--31.0) &	273&\\
   106	&	06 07 47.49 &	-06 23 36.4 &	46 &	99.7 &	1.4(0.8--2.3) &	54.6(54.0--55.3) &	3.5(2.1--5.4) &	5.7 &	30.5(30.1--31.1) &	274&\\
   107	&	06 07 47.59 &	-06 22 45.4 &	88 &	100.0 &	7.1(2.9) &	54.5(54.2--55.3) &	6.5(4.0--10.8) &	24.2 &	30.7(30.5--31.1) &	&\\
   108	&	06 07 47.64 &	-06 24 19.8 &	56 &	99.9 &	3.4(1.8--7.4) &	53.9(53.6--54.2) &	0.8(0.5--1.5) &	6.9 &	30.0(29.9--30.1) &	279&\\
   109	&	06 07 47.66 &	-06 23 58.6 &	18 &	59.5 &	3.0(fixed) &	53.3(53.1--53.6) &	0.1(0.0--0.5) &	2.9 &	29.4(29.4--29.6) &	277&\\
   110	&	06 07 47.67 &	-06 22 55.3 &	70 &	100.0 &	4.7(2.8--15.6) &	53.8(53.6--54.1) &	0.6(0.2--1.2) &	8.0 &	30.0(29.9--30.1) &	278&\\
   111	&	06 07 47.72 &	-06 23 11.9 &	14 &	69.7 &	3.0(fixed) &	53.5(53.0--53.7) &	1.8(0.7--4.3) &	1.9 &	29.5(29.2--29.7) &	283&\\
   112	&	06 07 47.77 &	-06 21 36.9 &	14 &	46.6 &	3.0(fixed) &	53.6(53.1--53.8) &	1.2(0.4--2.6) &	2.6 &	29.6(29.2--29.7) &	284&\\
   113	&	06 07 47.80 &	-06 24 09.6 &	6 &	86.3 &	3.0(fixed) &	52.9(52.1--53.4) &	0.4(0.0--4.5) &	0.9 &	29.0(28.7--29.4) &	286&\\
   114	&	06 07 47.81 &	-06 23 30.9 &	5 &	34.0 &	3.0(fixed) &	53.2(51.9--53.8) &	3.0(0.0--23.9) &	0.9 &	29.3(28.3--29.8) &	210&\\
   115	&	06 07 47.84 &	-06 21 45.1 &	15 &	78.7 &	3.0(fixed) &	53.6(53.2--53.8) &	1.4(0.6--2.7) &	2.5 &	29.6(29.3--29.7) &	287&\\
   116	&	06 07 47.87 &	-06 22 55.5 &	46 & 68.4 &	?\phd?($>$6.2)$^{\rm g}$ &	53.8(53.6--53.9) &	1.6(0.9--3.1) & 9.2 &	30.0(29.9--30.2) &	291& IRS~3NE\\
   117	&	06 07 47.88 &	-06 21 41.9 &	15 &	100.0 &	3.0(fixed) &	54.1(53.7--54.4) &	6.5(3.2--13.3) &	4.8 &	30.2(29.8--30.3) &	292&\\
   118	&	06 07 48.03 &	-06 22 30.8 &	194 &	98.8 &	2.5(1.8--3.7) &	55.1(54.9--55.4) &	5.2(4.1--6.7) &	40.6 &	31.1(31.0--31.3) &	296&e\\
   119	&	06 07 48.04 &	-06 22 39.3 &	244 &	51.3 &	2.2(1.7--3.0) &	54.6(54.4--54.7) &	0.9(0.7--1.1) &	22.8 &	30.6(30.5--30.7) &	299&f\\
   120	&	06 07 48.07 &	-06 21 22.6 &	54 &	48.9 &	?\phd?($>$5.6)$^{\rm g}$ &	54.6(54.4--54.7) &	5.8(3.8--10.4) &	44.0 &	30.8(30.7--31.0) &	&\\
   121	&	06 07 48.09 &	-06 23 33.0 &	19 &	97.7 &	3.0(fixed) &	54.0(53.6--54.3) &	7.9(3.0--17.3) &	3.8 &	30.1(29.7--30.3) &	&\\
   122	&	06 07 48.35 &	-06 23 22.4 &	8 &	95.0 &	3.0(fixed) &	53.3(52.2--53.6) &	2.5(0.0--8.5) &	1.1 &	29.3(28.4--29.5) &	&\\
   123	&	06 07 48.55 &	-06 21 32.2 &	9 &	92.8 &	3.0(fixed) &	53.2(52.5--53.5) &	0.4(0.0--1.9) &	1.4 &	29.2(28.9--29.5) &	317&\\
   124	&	06 07 48.64 &	-06 23 40.5 &	9 &	97.6 &	3.0(fixed) &	53.8(53.1--54.1) &	4.1(1.0--11.8) &	2.7 &	29.8(29.2--30.0) &	320&\\
   125	&	06 07 48.68 &	-06 24 06.3 &	128 &	100.0 &	3.2(2.1--6.0) &	54.6(54.3--54.9) &	2.0(1.4--2.9) &	25.8 &	30.7(30.6--30.8) &	325&\\
   126	&	06 07 48.69 &	-06 22 27.4 &	99 &	98.2 &	7.6(3.3--69.9) &	54.1(54.0--54.4) &	1.5(0.9--2.3) &	18.1 &	30.4(30.3--30.5) &	323&\\
   127	&	06 07 48.83 &	-06 22 21.8 &	14 &	55.9 &	3.0(fixed) &	53.2(52.5--53.6) &	0.8(0.0--3.1) &	1.2 &	29.2(28.8--29.5) &	328&\\
   128	&	06 07 49.03 &	-06 21 39.7 &	83 &	96.9 &	5.9(2.7--23.9) &	54.2(54.0--54.4) &	0.6(0.3--1.1) &	20.4 &	30.4(30.3--30.4) &	332&\\
   129	&	06 07 49.07 &	-06 22 17.7 &	8 &	75.9 &	3.0(fixed) &	52.4(51.2--52.7) &	0.0(0.0--0.0) &	0.3 &	28.4(28.4--28.4) &	334&\\
   130	&	06 07 49.07 &	-06 24 22.7 &	159 &	96.4 &	1.9(1.5--2.5) &	54.1(54.0--54.2) &	0.2(0.1--0.4) &	10.3 &	30.1(30.0--30.1) &	336&\\
   131	&	06 07 49.25 &	-06 22 48.1 &	9 &	74.9 &	3.0(fixed) &	53.3(52.8--53.6) &	2.0(0.4--4.9) &	1.3 &	29.4(28.8--29.5) &	&\\
   132	&	06 07 49.37 &	-06 23 54.3 &	8 &	68.5 &	3.0(fixed) &	53.6(52.7--54.0) &	2.8(0.2--13.5) &	2.0 &	29.6(28.8--30.0) &	&\\
   133	&	06 07 49.62 &	-06 23 22.8 &	11 &	78.3 &	3.0(fixed) &	53.5(53.0--53.8) &	2.6(0.9--7.0) &	1.8 &	29.6(29.1--29.8) &	345&\\
   134	&	06 07 49.65 &	-06 21 42.6 &	60 &	100.0 &	3.7(1.6--9.1) &	54.0(53.7--54.3) &	0.7(0.4--1.4) &	9.3 &	30.1(30.0--30.3) &	346&\\
   135	&	06 07 49.82 &	-06 21 14.0 &	29 &	67.2 &	1.4(0.7--3.1) &	53.9(53.4--54.7) &	1.4(0.7--2.7) &	2.4 &	29.8(29.6--30.5) &	349&\\
   136	&	06 07 49.93 &	-06 22 10.8 &	8 &	70.4 &	3.0(fixed) &	53.2(52.6--53.6) &	0.2(0.0--2.3) &	1.6 &	29.2(29.1--29.6) &	352&\\
   137	&	06 07 49.97 &	-06 23 20.1 &	23 &	97.5 &	1.0(0.4--2.2) &	54.4(53.7--55.7) &	2.8(1.3--5.4) &	2.5 &	30.3(29.7--31.3) &	353&\\
   138	&	06 07 50.06 &	-06 22 36.0 &	9 &	84.9 &	3.0(fixed) &	53.5(52.8--53.8) &	3.8(0.9--11.1) &	1.6 &	29.6(28.9--29.8) &	354&\\
   139	&	06 07 50.35 &	-06 23 46.9 &	10 &	89.2 &	3.0(fixed) &	53.4(52.9--53.7) &	2.4(0.0--8.8) &	1.4 &	29.5(28.5--29.7) &	355&\\
   140	&	06 07 50.43 &	-06 24 05.0 &	11 &	58.7 &	3.0(fixed) &	53.0(52.5--53.3) &	0.5(0.0--1.7) &	1.0 &	29.1(28.7--29.3) &	358&\\
   141	&	06 07 50.57 &	-06 24 01.9 &	7 &	67.3 &	3.0(fixed) &	52.8(51.4--53.3) &	0.9(0.0--5.6) &	0.5 &	28.8(28.4--29.0) &	361&\\
   142	&	06 07 50.58 &	-06 23 28.5 &	16 &	59.3 &	3.0(fixed) &	53.4(53.0--53.7) &	0.1(0.0--0.8) &	3.1 &	29.5(29.4--29.7) &	360&\\
   143	&	06 07 50.75 &	-06 23 23.1 &	6 &	96.9 &	3.0(fixed) &	52.8(52.2--53.3) &	0.0(0.0--0.0) &	0.9 &	28.9(28.9--28.9) &	362&\\
   144	&	06 07 51.00 &	-06 22 60.0 &	9 &	74.2 &	3.0(fixed) &	53.5(52.8--54.1) &	3.1(0.7--22.7) &	1.5 &	29.5(28.9--30.2) &	364&\\
   145	&	06 07 51.04 &	-06 22 24.3 &	7 &	53.1 &	3.0(fixed) &	53.6(52.3--54.0) &	2.1(0.0--16.2) &	2.2 &	29.6(28.8--30.0) &	366&\\
   146	&	06 07 51.39 &	-06 21 44.5 &	26 &	97.3 &	1.8(0.6--5.1) &	53.4(53.2--54.5) &	0.5(0.1--2.2) &	1.6 &	29.4(29.2--30.3) &	369&\\
   147	&	06 07 51.70 &	-06 23 18.1 &	79 &	87.1 &	2.7(1.6--5.8) &	55.0(54.7--55.4) &	1.7(1.1--2.6) &	61.8 &	31.1(30.9--31.3) &	372&\\
   148	&	06 07 51.74 &	-06 21 29.8 &	96 &	99.9 &	3.1(1.5--6.6) &	54.3(53.9--54.7) &	1.5(0.9--2.6) &	12.7 &	30.3(30.2--30.6) &	374&\\
   149	&	06 07 52.17 &	-06 21 17.7 &	68 &	90.0 &	2.1(1.2--3.2) &	53.6(53.5--53.9) &	0.2(0.0--0.4) &	4.1 &	29.6(29.5--29.7) &	378&\\
   150	&	06 07 52.25 &	-06 21 38.5 &	70 &	87.4 &	1.5(1.1--2.1) &	53.8(53.6--54.5) &	0.3(0.1--1.0) &	4.2 &	29.7(29.6--30.3) &	379&\\
   151	&	06 07 52.28 &	-06 21 18.2 &	69 &	35.0 &	2.1(1.2--3.6) &	53.7(53.5--53.9) &	0.2(0.1--0.5) &	4.3 &	29.7(29.6--29.8) &	380&\\
   152	&	06 07 52.30 &	-06 23 46.5 &	12 &	98.7 &	3.0(fixed) &	53.3(52.9--53.6) &	1.4(0.4--3.1) &	1.5 &	29.4(29.0--29.6) &	381&\\
   153	&	06 07 52.38 &	-06 22 57.4 &	10 &	62.7 &	3.0(fixed) &	53.4(53.0--53.7) &	0.0(0.0--0.8) &	3.5 &	29.5(29.4--29.7) &	382&\\
   154	&	06 07 52.91 &	-06 23 27.6 &	11 &	88.8 &	3.0(fixed) &	52.7(52.3--52.9) &	0.0(0.0--0.0) &	0.7 &	28.8(28.8--28.8) &	386&\\
\enddata
\tablecomments{
Parentheses indicate the 90\% confidence limits.}
\tablenotetext{a}{ Mean position error is 0.15$''$.}
\tablenotetext{b}{Variable probability obtained by Kolmogorov-Smirnov test to 0.5--10~keV band.}
\tablenotetext{c}{Observed flux in 0.5--10.0~keV band [$10^{-15}~{\rm ergs~s^{-1}~cm^{-2}}$].}
\tablenotetext{d}{Absorption-corrected luminosity in 0.5--10.0~keV band.}
\tablenotetext{e}{IR counterpart, source numbers are from Carpenter et al.\ (1997).}
\tablenotetext{f}{IR source names from Beckwith et al.\ (1976), Howard et al.\ (1994) and Capenter et al.\ (1997).}
\tablenotetext{g}{The best-fit values of $kT$ not determined to be below 20~keV, hence
 we assume the temperatures are 20~keV for estimation of the other parameters.}

\end{deluxetable}

\end{document}